\documentclass[conference,10pt]{IEEEtran}
\usepackage[dvips]{color}
\usepackage{epsf}
\usepackage{times}
\usepackage{epsfig}
\usepackage{graphicx}

\usepackage{amsmath}
\usepackage{amssymb}
\usepackage{amsxtra}
\usepackage{amsthm}
\usepackage{bbm}

\usepackage{here}
\usepackage{rawfonts}
\usepackage{times}
\usepackage{url}
\usepackage{cite}
\usepackage{comment}
\usepackage[utf8]{inputenc}
\usepackage{caption}
\usepackage{subcaption}

\usepackage{pstricks}
\usepackage{algorithm}
\usepackage{algpseudocode}
\usepackage{lipsum}
\usepackage{mathtools}

\makeatletter 

\IEEEoverridecommandlockouts

\begin{document}\vspace{-0.2cm}
\title{\huge Reflections over the Sea: Reconfigurable Intelligent Surface for Maritime Self-Powered Communications}  
\vspace{-0.8cm}
\author{ 
	\IEEEauthorblockN{ Qianqian Zhang$^\ast$,  Long Wang$^\dagger$, Ben Wu$^\ast$, and Jia Mi$^\S$ } 	
	\IEEEauthorblockA{\small 
		$^\ast$Department of Electrical and Computer Engineering, Rowan University, NJ, USA,
		Emails: \url{{zhangqia,wub}@rowan.edu}. \\
		$^\dagger$Department of Mechanical Engineering, Stevens Institute of Technology, NJ, USA Emails: \url{lwang4@stevens.edu}. \\
		$^\S$Department of Civil, Environmental and Ocean Engineering, Stevens Institute of Technology, NJ, USA Emails: \url{jmi5@stevens.edu}.
		 \vspace{-0.6cm}
	}
} 
\maketitle

\setlength{\columnsep}{0.55cm}
\begin{abstract}
Maritime communication is becoming a vital component of 6G networks, driven by the rapid expansion of the maritime economy. However, existing technologies face critical challenges in signal coverage, availability, and robustness, especially under harsh sea conditions. This paper proposes a novel framework for the maritime Internet-of-Things (IoT) communications that  leverages the reconfigurable intelligent surface (RIS) mounted on offshore infrastructures, such as wind turbines,  to enhance coverage and reliability. To capture dynamic maritime environment,  a near-ocean-surface channel model is developed considering the impact of sea waves. In addition, a wave energy harvesting (EH) system is designed to self-power IoT sensors for data acquisition, processing, and transmission. To support real-time adaptation, channel state information is continuously measured to optimize RIS reflection parameters and maximize multi-user communication rates. 
Simulation results show that the proposed system significantly improves IoT communication performance by over $20\%$, under harsh sea conditions. 

 
\end{abstract}

\IEEEpeerreviewmaketitle

\vspace{-0.2cm}
\section{Introduction}

With the advent of sixth generation (6G) networks, wireless networks are expected to extend beyond terrestrial boundaries to include aerial, space, and maritime domains—collectively referred to as non-terrestrial networks (NTNs). 
Among these, maritime communication is emerging as a critical component, driven by the rapid growth of  maritime economy, including offshore infrastructure,  autonomous vessels, smart ports, and ocean sensing networks, with maritime Internet-of-Things (IoT) devices increasingly deployed to support diverse applications \cite{huo2020cellular}. 
For instance, offshore energy operations and environmental monitoring require robust  IoT connectivity in remote and harsh ocean environments, and autonomous vessels and smart ports rely on reliable communication for safe navigation and real-time coordination. These growing demands highlight the need for resilient and efficient maritime communication technologies as a foundational pillar of  6G networks.


Compared with terrestrial and other NTN systems, maritime communication faces distinct challenges arising from the dynamic ocean environment, sparse infrastructure, remote locations, and harsh environment conditions \cite{wang2018wireless}.  
Maritime channels, dominated by line-of-sight (LoS) and sea-surface reflection paths, exhibit sparse scattering and experience significant signal attenuation when LoS links are obstructed by vessels or ocean waves.   
Sea-induced motion also causes  continuous fluctuations in antenna height and orientation, leading to link mismatches and variations in received signal strength \cite{alqurashi2022maritime}. 
Moreover, the remoteness of maritime operations  results in long transmission distances,  which introduce additional challenges related to latency, coverage, and power supply.  

Current maritime data technologies are facing different limitations   \cite{alqurashi2022maritime}. 
To ensure a wide coverage, conventional maritime communication systems rely on satellites. However, the scarce bandwidth, as well as the high service and power costs,  makes them unsuitable to scale for emerging applications requiring higher data rates and faster response. 
Cellular-based maritime IoT systems, such as WiMAX and LTE, have been investigated  in near-shore scenarios \cite{huo2020cellular}, but their effectiveness diminishes in offshore environments.   
Aerial platforms including unmanned aerial vehicles (UAVs) and high altitude platforms (HAPs) \cite{wang2021hybrid} offer flexible  deployment, but are limited by the intermittent coverage and low reliability under adverse weather conditions, rendering them unsuitable for real-time, mission-critical maritime applications such as emergency response. 



To address the aforementioned challenges,  this paper proposes a novel solution to deploy the reconfigurable reflection surface (RIS) onto offshore infrastructures, such as wind turbines and drilling platforms, to enhance the maritime communication environment for  IoT devices. 
The RIS reflects the wireless signals from  maritime IoT nodes toward a local data center to improve the communication performance, reliability, and coverage.  
To adapt for maritime conditions, a near-ocean-surface channel model under the impact of sea waves is developed. 
Additionally, a wave energy harvesting (EH) system is designed to power the IoT sensors for data acquisition, processing, and transmission. 
To maximize the multi-user (MU) IoT communication rates, the channel state information (CSI) is measured in a real-time manner  to optimize the RIS reflection parameter. 
To the best of our knowledge, this is the first paper that proposes the deployment of RIS for maritime communications.  
Moreover, the cross-disciplinary integration of mechanical design and communication network enhances the resilience of the proposed solution, thus supporting broader applicability across diverse maritime scenarios.


The rest of this paper is organized as follows. 
Section II presents the self-powered communication system, with the maritime channel model, RIS-assisted communication model, and EH model. 
The real-time channel estimation and reflection optimizations are presented in Section III. 
Simulation results are shown in Section IV. Conclusions are drawn in Section V.

\begin{figure*}[!t]
	\begin{center}\vspace{-0.0cm}
		\begin{subfigure}{.32\textwidth}
			\centering
			\includegraphics[width=5.4cm]{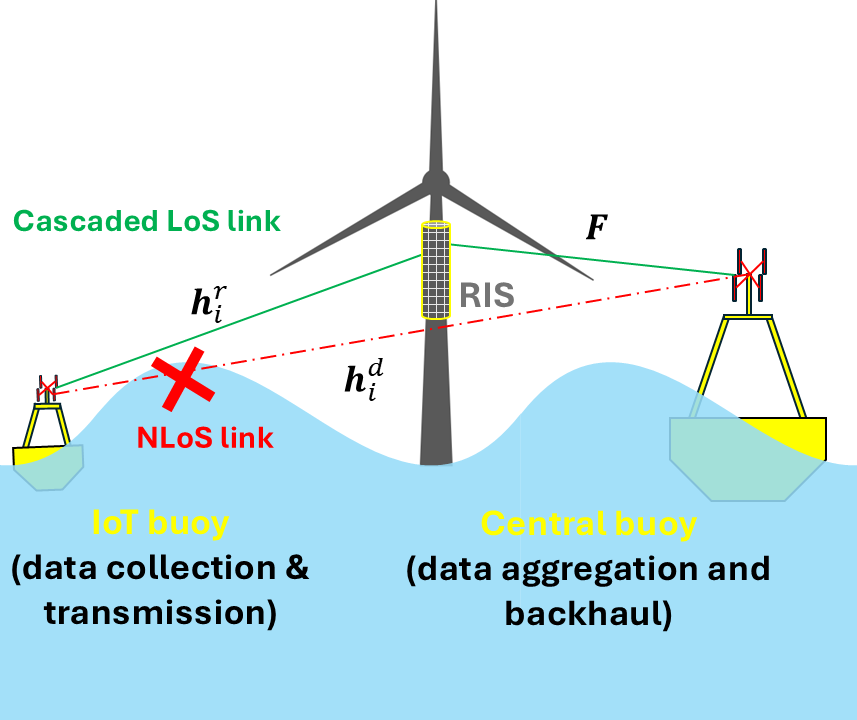}
			\caption{\label{fig_11} }
		\end{subfigure}
		\begin{subfigure}{.33\textwidth}
			\centering
			\includegraphics[width=6.1cm]{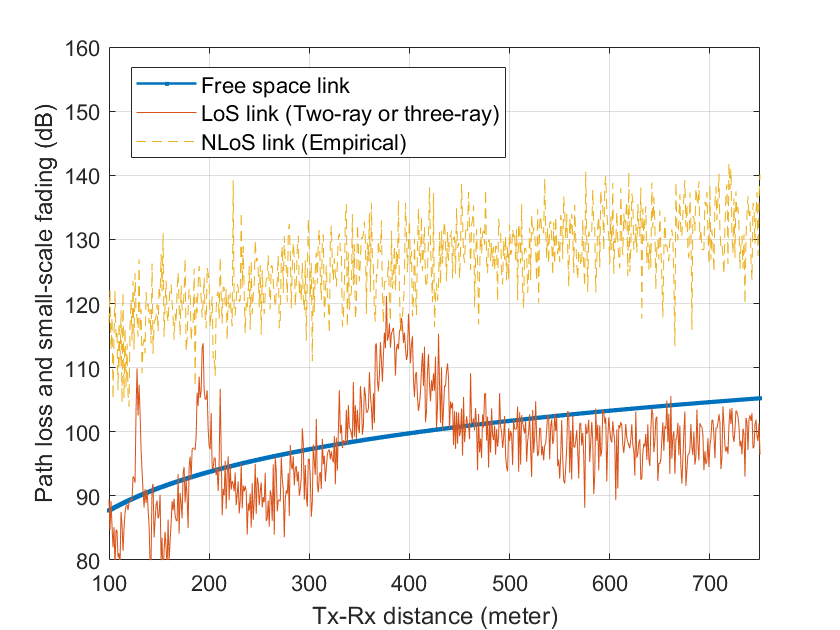}
			\caption{\label{fig13} }
		\end{subfigure}
		\begin{subfigure}{.31\textwidth}
			\centering
			\includegraphics[width=5.4cm]{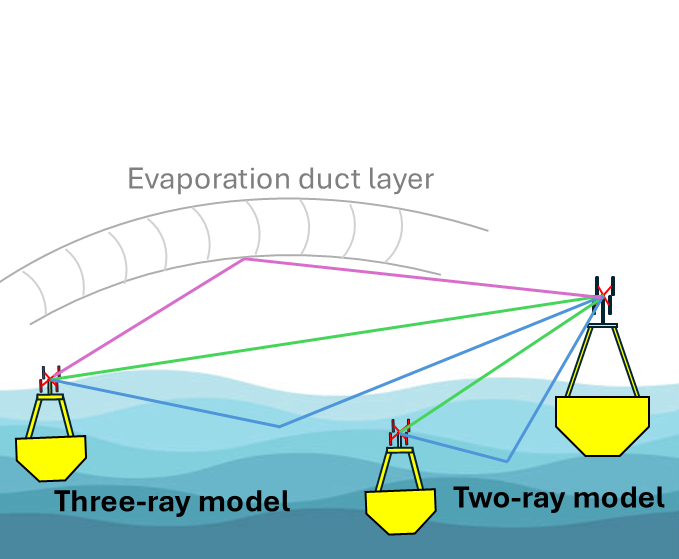}
			\caption{\label{fig12}  }
		\end{subfigure}
		\vspace{-0.3cm}
		\caption{\small{\label{fig1} (a) RIS-aided maritime IoT communication system. (b)   Path loss and small-scale fading for free space, LoS, and NLoS links, given $f_c=5.8$ GHz, $h_t=2$ meters, $h_r=5$ meters, and $h_e=50$ meters. (c) Two-ray and three-ray LoS channel models. }
		}
	\end{center}
	\vspace{-0.7cm}
\end{figure*}

\vspace{-0.2cm}

\section{System Model} 


Consider a maritime sensor system where each IoT device is mounted onto a buoy and deployed within an offshore wind farm or near drilling platforms to support the safety and environmental tasks. 
As shown in  Fig. \ref{fig_11},  each IoT collects task-specific environmental data and transmit it to a central node  located on a large buoy. 
The center buoy aggregates data from surrounding  buoys, performs local processing, and issues anomaly alerts when necessary. 
However, continuous wave-induced motion  causes significant fluctuations in buoy height, resulting in channel variation and antenna misalignment. 
Under harsh weather conditions, sea waves can block line-of-sight (LoS) links, which severely degrades signal strength and communication reliability. 
To address this issue, an RIS is deployed on nearby offshore infrastructure. 
The elevated height  enables the RIS to maintain LoS paths with both transmitter (Tx) and receiver (Rx), enabling efficient signal reflection and enhancing link robustness, while imposing minimal operational burden on the offshore platform.




\vspace{-0.1cm} 
\subsection{Maritime Channel Model} \label{sec_channelModel}
Given the transmit power $P_t$ in dB, the receive signal power is expressed as
$P_r = P_t + G_t - L(\boldsymbol{t},\boldsymbol{r})+ G_r$, 
where $G_t$ and $G_r$ denote the Tx's and Rx's antenna gains,  $\boldsymbol{t}=(x_t,y_t)$ and $\boldsymbol{r}=(x_r,y_r)$ represent the antenna locations for Tx and Rx, respectively, and $ L$ denotes the path loss whose value depends on the LoS and non-line-of-sight (NLoS) link conditions.  

Initially, we consider a calm ocean surface with limited wave height, where the LoS link exists and the sea acts as a large reflective surface. 
Under these conditions,  the near-ocean propagation channel can be accurately modeled using the two-ray model, as shown in Fig. \ref{fig12}. 
However,  as the propagation distance $d=|\boldsymbol{t}-\boldsymbol{r}|$ becomes larger,  a third propagation  path originating from the evaporation duct layer becomes non-negligible. 
In this case, the three-ray model provides a more precise characterization of the channel response \cite{wang2018wireless}. 
Accordingly, the LoS path loss in dB is expressed as
\begin{multline}  \label{equ_pl_los}  \vspace{-0.1cm}
	L_{LoS}(h_t,h_r,d,h_e)   =  \\
	\begin{cases}
		-20 \log_{10}\left[  \frac{\lambda}{2\pi d}    \sin\left(\frac{2 \pi h_t h_r}{\lambda d}\right)  \right]  + \xi_{LoS} & \text{if } d \le \frac{4h_t h_r}{\lambda},\\
		-20 \log_{10} \left[ \frac{\lambda}{2\pi d} (1+\Delta)  \right] + \xi_{LoS} & \text{if }d > \frac{4h_t h_r}{\lambda},
	\end{cases}	 
\end{multline}  
where $\lambda$ is the carrier wavelength, $h_t$ and $h_r$ are the Tx and Rx antenna heights relative to the sea surface (defined later in Section \ref{sec_wave}), and $h_e$ is the effective height of the evaporation duct. 
Here, $\Delta = 2\sin\left(\frac{2 \pi h_t h_r}{\lambda d}\right)\sin\left[ \frac{2 \pi (h_e-h_t) (h_e-h_r)}{\lambda d}\right]$ characterizes the three-ray  model, 
and $\xi \sim \mathcal{N}(0,\sigma^2_{LoS})$ represents the small-scale fading for LoS links  \cite{reyes20145}. 

As wave height increases,  the LoS link may eventually be obstructed  by ocean water. 
In such cases, a scattering channel model, based on the empirical measurement \cite{reyes2011buoy}, is employed to characterize the NLoS link. 
The NLoS path loss model accounts for multi-path reflections and fading effect caused by the rough and dynamic ocean surface, given as: 
\begin{equation} \vspace{-0.1cm}
	L_{NLoS}(d)   =   K + 10 \alpha \log_{10}(d/d_0) + \xi_{NLoS}, 
\end{equation}  
where $d_0$ is a reference distance in meters, while $K$, $\alpha$, and $\xi_{NLoS}\sim \mathcal{N}(0,\sigma^2_{NLoS})$ are all in dB units given in \cite{reyes2011buoy}. 
In Fig. \ref{fig13}, the maritime path loss for both LoS and NLos links with small-scale fading are shown with respect to the Tx-Rx distance $d$.  For comparison, the free-space path loss is included as a baseline, given by $L_{fs} =-147.55 + 20 \log_{10}(f_c)+20 \log_{10}(d)$, where $f_c$ is the carrier frequency in Hz. 


\begin{figure*}[!t]
	\begin{center}\vspace{-0.0cm}
		\begin{subfigure}{.31\textwidth}
			\centering
			\includegraphics[width=5.2cm]{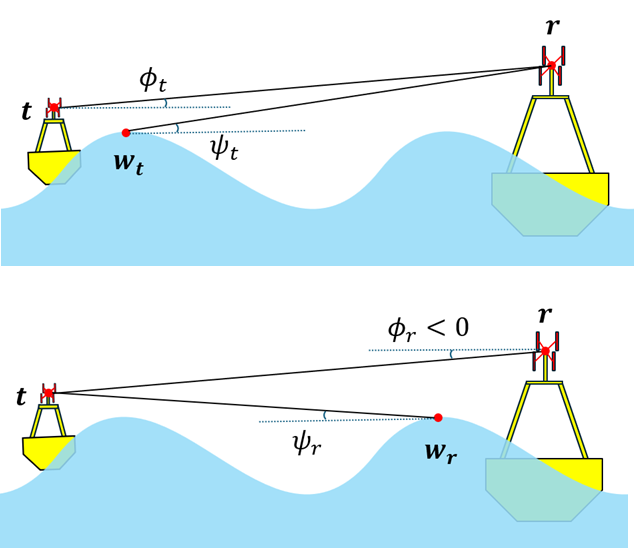}
			\caption{\label{fig_21} }
		\end{subfigure}
		\begin{subfigure}{.33\textwidth}
			\centering
			\includegraphics[width=5.8cm]{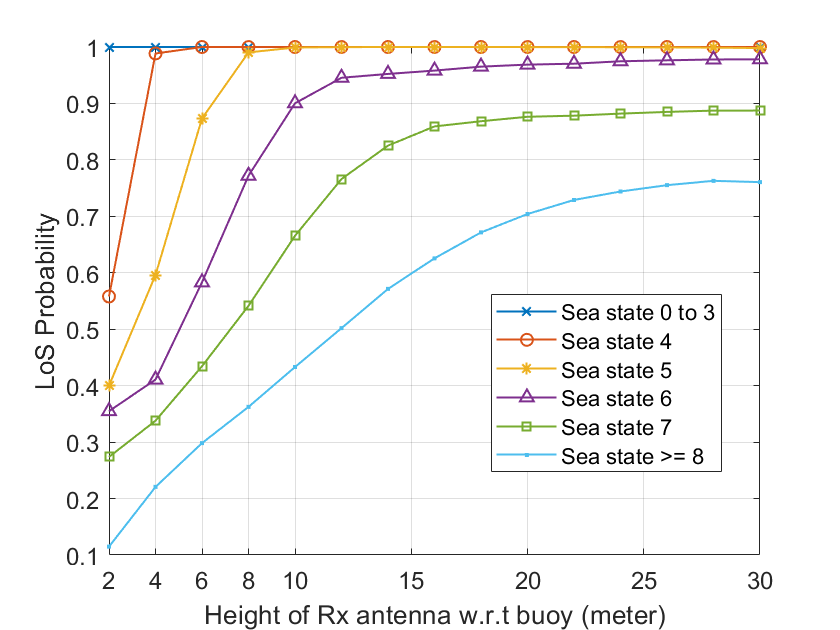}
			\caption{\label{los_hr} }
		\end{subfigure}
		\begin{subfigure}{.33\textwidth}
			\centering
			\includegraphics[width=6.4cm]{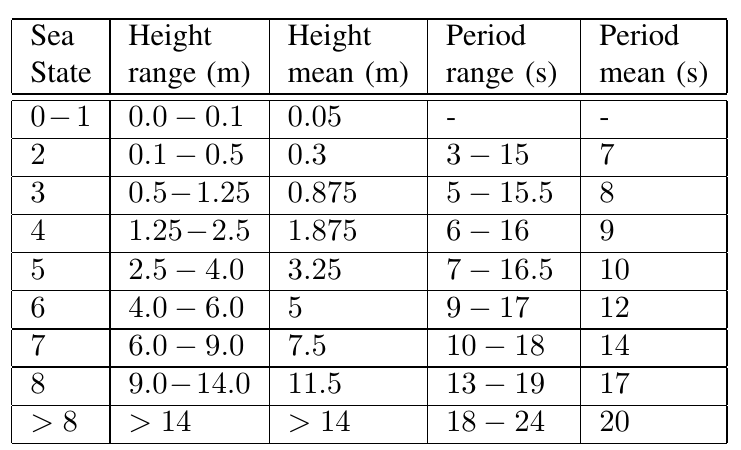}
			\caption{\label{table}  }
		\end{subfigure}
		\vspace{-0.3cm}
		\caption{\small{\label{fig2} (a) Impact of wave to Tx-Rx LoS link. (b)  LoS probability for the Tx-Rx direct link, under different sea states, given various height of the Rx antenna. (c) Wave height and period under different sea states of open ocean in North Atlantic. }
		}
	\end{center}
	\vspace{-0.7cm}
\end{figure*}

\vspace{-0.1cm}
\subsection{Impact of Sea Wave}\label{sec_wave}

Given that  both the Tx and the Rx are mounted onto buoys floating on the sea surface, each  experiences periodic motion due to ocean waves. 
In the self-powered maritime communication,  wave-induced dynamics introduce three primary effects: 
First, as shown in Eq. (\ref{equ_pl_los}), fluctuations change the Tx and Rx antenna heights $h_t$ and $h_r$ which in turn affect  the LoS path loss;  
Second, high waves may obstruct the LoS path, causing severe  signal attenuation;  
Third, ocean motion impacts the amount of energy harvested from the waves, which will be discussed in Section \ref{sec_eneHarvest}.   
Among the six degrees of freedom (DoF) of the buoy's heave motion, the height variation of the antenna becomes the most critical factor influencing communication performance. 

Following the typical sine wave movement pattern \cite{wang2011effective}, 
the vertical position of the Rx antenna in Eq. (\ref{equ_pl_los}) can be given by  
\begin{equation}\small
	h_r (t)= a \sin\left[2\pi \frac{\text{mod}(d_r(t), l)}{l} + 2\pi \frac{\text{mod}(t, \mathcal{T})}{\mathcal{T}}\right] + h_r^0, 
\end{equation}\normalsize
where  $d_r(t) = \sqrt{[x_r(t)- x_0]^2 + [y_r(t)-y_0]^2}$ is the distance from the dominant wave source to the Rx at time $t$, $a$, $l$, and $\mathcal{T}$ are the amplitude, length, and period of the wave height, respectively,  and $h_r^0$ is the antenna height relative to the buoy. The antenna height of Tx $h_t(t)$ will follow a similar pattern. 

As shown in Fig. \ref{fig_21},  signal propagation between Tx and Rx is influenced by peak points along the sea surface. 
Let $\boldsymbol{w}_t$ and $\boldsymbol{w}_r$ be the locations of the closest wave peaks to Tx and Rx, respectively.   
The line-of-sight (LoS) link is maintained if the direct path between Tx and Rx lies above the relevant wave peak;   otherwise, the link experiences NLoS propagation. 

The elevation angle from Tx to Rx is defined  as 
\begin{equation}
	\phi_t(t)=\arctan \left( \frac{h_r(t)-h_t(t)}{|\boldsymbol{r} - \boldsymbol{t}|} \right). 
\end{equation}
Similarly, the angle between the horizontal plane and the line connecting the closest wave peak of Tx to Rx  is: 
\begin{equation}\label{equ_beta}
	\psi_t(t)=\arctan \left( \frac{h_r(t) -a}{|\boldsymbol{r}(t) - \boldsymbol{w}_t(t)|} \right),  
\end{equation}
and the location $\boldsymbol{w}_t(t)$ of the closest wave peak to Tx is  \cite{wang2011effective}
\begin{equation}\label{equ_peak}
	\boldsymbol{w}_t(t)  =  
	\begin{cases}
		 \boldsymbol{t}(t) + l [1- \frac{a-h_t(t)}{4a}]   & \text{\small if Tx is moving downwards}, \\
		\boldsymbol{t}(t) + \frac{l[a-h_t(t)]}{4a} & \text{\small if Tx is moving upwards}. 
	\end{cases}	
\end{equation} 
When $\phi_t(t) \le \psi_t(t)$, the closest wave peak of Tx does not block the LoS link. 
Similarly, 
the elevation angle from Rx to Tx  is $\phi_r(t)= -\phi_t(t)$, and the corresponding blocking angle  $\psi_r(t)$ can be defined analogously to Eqs. (\ref{equ_beta}) and (\ref{equ_peak}). 

Consequently, a LoS link exists only if both $\phi_t(t) \le \psi_t(t)$ and $\phi_r(t) \le \psi_r(t)$ are satisfied. 
As shown in Fig. \ref{los_hr}, under various sea states of open ocean in North Atlantic, the LoS probability increases with the Rx antenna height $h_r^0$, assuming a fixed Tx height of $h^0_t=2$ meters. 
For sea states below level four where wave height stays mostly under two meters, the LoS link can be consistently maintained even with a low Rx antenna height of two meters.   
In higher sea states, severe wave dynamics cause frequent link blockage. 
Even Rx antennas at $30$ meters may not sustain reliable LoS, while in reality, the antenna height on a buoy  stays no higher than $8$ meters. 



\begin{figure}[!t]\vspace{-0.3 cm}
	\begin{center}  
		\includegraphics[width=5.2cm]{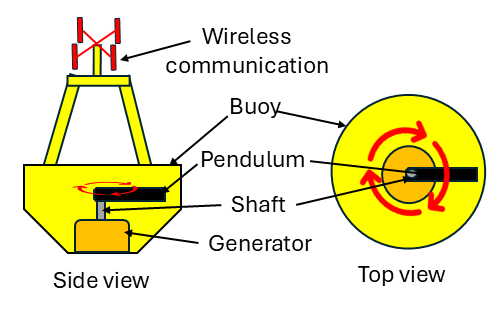} \vspace{-0.2 cm}
		\caption{\label{buoy}\small EH design with a wave energy converter. }  
	\end{center}\vspace{-0.7cm}  
\end{figure}

\vspace{-0.1cm}

\subsection{Energy Harvesting at Buoy}\label{sec_eneHarvest}

While ocean waves can significantly attenuate wireless signals, they also offer a valuable energy source for self-powered systems.  
To reduce maintenance costs, each buoy is equipped with an EH system that powers maritime IoT devices for data collection, processing, and wireless transmissions.    
As shown in Fig. \ref{buoy},  the proposed wave energy converter (WEC) employs  a pendulum-based design to harness the hydrokinetic energy of ocean waves \cite{mi2025multi}.  
An electrical generator, mounted at the base of the buoy, is mechanically coupled to a pendulum through a central shaft. 
Wave-induced oscillations cause the pendulum to swing and rotate, converting mechanical motion into electrical energy. 
This harvested energy powers the wireless communication module, offering a sustainable and efficient solution for marine IoT deployments.

Given the wave amplitude $a$ and period $\mathcal{T}$, the available power per unit wavelength is 
\begin{equation} \vspace{-0.2cm}
	P_{\text{wave}} (a,\mathcal{T}) = \frac{\rho g^2}{64 \pi} a^2 \mathcal{T},
\end{equation}  
where $\rho $ is the seawater density  and $g$ is the gravitational acceleration. 
For a point absorber buoy, the harvested  power  depends on the effective interaction width $W$ between the buoy and the incident  wave, and is given by:
\begin{equation} \label{equ_power}
	P_e (a,\mathcal{T},W) = \eta \cdot    P_{\text{wave}} \cdot W, 
\end{equation}
where $\eta  = \eta_{\text{pto}} \cdot \eta_{\text{conv}} \cdot \gamma_\text{cwr}$ is the  wave-to-wire conversion efficiency.   
Here, $\eta_{\text{pto}}$ and  $\eta_{\text{conv}}$ represent the energy conversion efficiency of the power takeoff (PTO) system and onboard electronics, respectively, and  $\gamma_\text{cwr}$ is the capture width ratio of captured power relative to the wave power incident on the device’s physical width \cite{babarit2015database}.  
Finally, given the operational power consumption $P_0$ of the buoy and IoT device, the available power for wireless transmission  will be $P_t\le P_e -P_0$.

\vspace{-0.1cm}

\subsection{RIS-aided Communication System}

Consider a block-fading MU model in the maritime communications, where $I$ IoT devices (i.e.,  Txs) transmit uplink signals to the center buoy (i.e., Rx)  with the assistance of an RIS  using the same time and frequency. 
The center buoy and RIS are equipped with $M$ antennas and $N$ reflective elements, respectively, while each IoT device has a single antenna.  
The direct channel  from  IoT $i$ to the center buoy is denoted by $\boldsymbol{h}_i^d \in \mathcal{C}^{1\times M}$,  with each link following the channel model defined in Section \ref{sec_channelModel}. 
The channels from IoT $i$ to RIS and from RIS to the center buoy are denoted by
$\boldsymbol{h}_i^r \in \mathcal{C}^{1\times  N}$ and 
$\boldsymbol{F} \in \mathcal{C}^{N\times M}$, respectively.  
Since offshore infrastructures, such as wind turbines, rise significantly above ocean surface, the RIS mounted on these structures can maintain the LoS links with both the IoT buoy to the center buoy.  
 
Let  $\boldsymbol{Q} = \text{diag}(\boldsymbol{q})$ be the RIS reflection coefficient, where $\boldsymbol{q}=[e^{j\theta_1},\cdots, e^{j\theta_N}] \in \mathcal{C}^{1\times N}$ and  $\theta_n$ is the phase shift of each RIS element \cite{zhang2019reflections}.  
Then, the combined channel response from IoT $i$ to the center buoy through RIS is denoted by 
\begin{equation} \vspace{-0.1cm}
	\boldsymbol{h}_i^d + \boldsymbol{h}_i^r \boldsymbol{Q} \boldsymbol{F} = \boldsymbol{h}_i^d + \boldsymbol{q} \boldsymbol{G}_i \in \mathcal{C}^{1\times M},
\end{equation}
where $\boldsymbol{G}_i =  \text{diag}(\boldsymbol{h}_i^r)\boldsymbol{F} \in \mathcal{C}^{N\times M}$ is the cascaded IoT-RIS-Rx link without any phase shift.  
Denote $s_i$ to be the transmitted signal from IoT node $i$.    
Then, the received signal at the center buoy can be expressed as:
\begin{equation} \vspace{-0.2cm}
	\boldsymbol{y} =  \sum_{i=1}^I (\boldsymbol{h}_i^d + \boldsymbol{q} \boldsymbol{G}_i) s_i + \boldsymbol{z},
\end{equation}
where  $\boldsymbol{z} \sim \mathcal{CN}(\boldsymbol{0},\sigma^2 \boldsymbol{I}_M)$ is the receiver noise vector.  

Consequently, with a bandwidth of $\beta$ , the sum capacity of the RIS-aided MU-MIMO system can be given by
\begin{equation}
	C= \beta \log_2 \det \left[ 1 + \frac{\sum_{i=1}^I ||(\boldsymbol{h}_i^d + \boldsymbol{q } \boldsymbol{G}_i)s_i||^2 }{\sigma^2} \right]. 
\end{equation}

\vspace{-0.1cm}

\subsection{Problem Formulation} 
The objective of  maritime IoT communications is to optimize the reflection coefficient of the RIS, so that the total uplink transmission of all IoT devices can be maximized,  under the real-time power constraint of each buoy, i.e.:   
\begin{subequations}\label{equs_Opt}  
	\begin{align}
		\max_{ \boldsymbol{q}=[e^{j\theta_1},\cdots, e^{j\theta_N}]  } \quad &     \sum_{i=1}^I ||(\boldsymbol{h}_i^d + \boldsymbol{q} \boldsymbol{G}_i)s_i||^2   \label{equ_obj}\\ 
		\textrm{s. t.} \quad  
		& |s_i|^2 \le \min\{P_{t,i}, P_{\text{max}} \}  ,  \forall i=1,\cdots, I  \label{cons_power}\\  
		& \theta_n \in [0,2\pi),  \forall n = 1,\cdots, N \label{cons_ris}
	\end{align}
\end{subequations}
where (\ref{cons_power}) limits the transmit power of each IoT buoy $i$ to not exceed the available onboard  power $P_{t,i}$ or the maximum transmit power $ P_{\text{max}}$,  and (\ref{cons_ris}) enforces the  reflection constraints at the RIS. 
Problem (\ref{equs_Opt}) is quadratically constrained quadratic program (QCQP) optimization, which can be challenging to solve, due to the dynamic nature of CSI, i.e., $\boldsymbol{h}_i^d$ and $\boldsymbol{G}_i$, for each IoT device $i$ in the maritime environment.  
Meanwhile, the objective function (\ref{equ_obj}) is non-convex with respect to $\boldsymbol{q}$, and the sparsity of maritime wireless channels may make the theoretical solution infeasible.   
To address these challenges, we will propose a practical solution in the next section to maximizes the capacity of RIS-aided maritime IoT communications. 


\section{Solution}\label{SecLearning}

Due to the ocean waves, the positions of both Txs and Rx vary continuously, resulting in time-varying CSI.   
For practical implementation,  we consider the block-fading channel to have a constant CSI during each coherence time. 
At the beginning of a coherence interval, pilot signals are transmitted from each IoT device to the central buoy for channel estimation. 
Based on the estimated result,  optimization is then applied to maximize the sum-rate of RIS-aided maritime communications.

\vspace{-0.1cm}

\subsection{Channel Estimation}  
At the beginning of each interval, the $i$-th IoT transmits its orthogonal pilot sequence with length $T$, i.e., $\boldsymbol{s}_i =[s_{i,1},\cdots,$ $s_{i,T}] \in \mathcal{C}^{1\times T}$ with $\boldsymbol{s}_i\boldsymbol{s}^H_i=P_{t,i} T$   and $\boldsymbol{s}_i\boldsymbol{s}^H_j=0$ if $i\ne j$, $\forall i,j$.  
Thus, the received pilot signal at the center buoy is
\begin{equation} \vspace{-0.1cm}
	\boldsymbol{Y} = \sum_{i=1}^{I} \boldsymbol{s}^H_i (\boldsymbol{h}_i^d + \boldsymbol{q} \boldsymbol{G}_i )  + \boldsymbol{Z},
\end{equation}
where $\boldsymbol{Z} \in \mathcal{C}^{T\times M}$ is the receiver noise matrix. 

Next, the channel estimation will be divided into two stages to evaluate the direct IoT-Rx channel and the cascaded IoT-RIS-Rx channel, respectively. 
In particular,  the direct channel estimation consists of two sub-frames, while the cascaded channel estimation includes $B \in \mathcal{N}^+$ sub-frames \cite{chen2023channel}. 

\subsubsection{Direct Channel Estimation}
In the first sub-frame, the RIS parameter is set as $\boldsymbol{q}_0 = [e^{j\theta_{1,0}},\cdots, e^{j\theta_{N,0}}]$ with the received signal denoted by $\boldsymbol{Y}_0$. In the second sub-frame, the reflection parameter is  $\boldsymbol{q}_1 = [e^{j\theta_{1,0}+\pi},\cdots, e^{j\theta_{N,0}+\pi}]=- \boldsymbol{q}_0$ with the received signal  $\boldsymbol{Y}_1$.  
Next, by adding $\boldsymbol{Y}_0$ and $\boldsymbol{Y}_1$, the cascaded IoT-RIS-Rx channel will be canceled out, i.e., 
\begin{multline}\vspace{-0.2cm} 
	\boldsymbol{Y}_0 + \boldsymbol{Y}_1 = 
	\sum_{i=1}^{I}  2 \boldsymbol{s}^H_i \boldsymbol{h}_i^d  + \sum_{i=1}^{I} \boldsymbol{s}^H_i (\boldsymbol{q}_0+\boldsymbol{q}_1) \boldsymbol{G}_i    + \boldsymbol{Z}_0 + \boldsymbol{Z}_1 \\
	= \sum_{i=1}^{I}  2\boldsymbol{s}^H_i  \boldsymbol{h}_i^d + \boldsymbol{Z}_0 + \boldsymbol{Z}_1 = 2 \boldsymbol{S} \boldsymbol{H}^H_d  + \boldsymbol{Z}_0 + \boldsymbol{Z}_1, 
\end{multline} 
where $\boldsymbol{H}_d=[(\boldsymbol{h}_1^d)^H,\cdots, (\boldsymbol{h}_I^d)^H] \in \mathcal{C}^{M\times I}$ and $\boldsymbol{S} =[\boldsymbol{s}^H_1,$ $\cdots,$ $ \boldsymbol{s}^H_I] \in \mathcal{C}^{T\times I} $. 
Then, the direct IoT-Rx channel can be measured using the least square (LS) estimator as
\begin{equation}\label{equ_estDCSI}
	\hat{\boldsymbol{H}}^H_d= \frac{1}{2}  (\boldsymbol{S}^H \boldsymbol{S})^{-1} \boldsymbol{S}^H (\boldsymbol{Y}_0 + \boldsymbol{Y}_1). 
\end{equation} 

\subsubsection{Cascade Channel Estimation}
Given the direct channel (\ref{equ_estDCSI}), the cascade channel through the RIS is estimated, using  pilot signals over additional $B$ sub-frames.  
Then, by subtracting the direct channel and applying the orthogonal pilot signal $\boldsymbol{s}_i$ to the received signal, we have that for the $b$-th sub-frame:   
\begin{equation} \label{equ_recPilot}
	\boldsymbol{u}_{i,b} \overset{\triangle}{=} \boldsymbol{s}_i (\boldsymbol{Y}_b - \boldsymbol{S}\hat{\boldsymbol{H}}^H_d) = \boldsymbol{q}_b \boldsymbol{G}_i + \tilde{\boldsymbol{z}}_{i,b},
\end{equation}
where $\tilde{\boldsymbol{z}}_{i,b} = \boldsymbol{s}_i \boldsymbol{Z}_b \in \mathcal{C}^{1\times M}$.   
Next, based on the pilot transmissions for  $B$ sub-frames, we define $\boldsymbol{U}_i = [\boldsymbol{u}^H_{i,1},\cdots,$ $\boldsymbol{u}^H_{i,B}] \in \mathcal{C}^{M\times B}$,  
$\tilde{\boldsymbol{Q}} = [\boldsymbol{q}^H_{1},$ $ \cdots, $ $ \boldsymbol{q}^H_{B} ]  \in \mathcal{C}^{N \times B}$, and $\tilde{\boldsymbol{Z}}_i=$ $[\tilde{\boldsymbol{z}}^H_{i,1}, \cdots, \tilde{\boldsymbol{z}}^H_{i,B}] \in \mathcal{C}^{M\times B}$. 
Then,  Eq. (\ref{equ_recPilot}) can be rewritten into the matrix form as
\begin{equation}
	\boldsymbol{U}_i =  \boldsymbol{G}^H_i \tilde{\boldsymbol{Q}} + \tilde{\boldsymbol{Z}}.
\end{equation}
Finally,  using LS estimator, the cascade channel from IoT $i$ through RIS to Rx without any phase shift is estimated as: 
\begin{equation}\label{equ_estCCSI}
	\hat{\boldsymbol{G}}^H_i = \boldsymbol{U}_i \tilde{\boldsymbol{Q}}^H   (\tilde{\boldsymbol{Q}} \tilde{\boldsymbol{Q}}^H)^{-1}. 
\end{equation}
It is worth noting that in (\ref{equ_estDCSI}) and (\ref{equ_estCCSI}),  LS estimator requires  both $\tilde{\boldsymbol{Q}}$ and $\boldsymbol{S}$ to be full-rank, i.e., $B\ge N$ and $T \ge I$. Therefore, the number of subframes and the pilot signal length must be sufficiently large to ensure reliable channel estimation.

\begin{algorithm}[t] 
	\caption{RIS-aided maritime IoT communications }
	\begin{algorithmic}
		\State \textbf{Channel estimation}: At the beginning of  $B+2$ sub-frames: \\
		(1) set the RIS parameter to be $\boldsymbol{q}_0,\boldsymbol{q}_1$, and $\tilde{\boldsymbol{Q}}$ sequentially;\\
		(2) each IoT device $i$ sends pilot signal $\boldsymbol{s}_i$ for  $B+2$ times; \\
		(3) the estimated CSI is $\hat{\boldsymbol{H}}_d$ in   (\ref{equ_estDCSI}) and  $\hat{\boldsymbol{G}}^H_i$ in (\ref{equ_estCCSI}); \\	
		\textbf{RIS optimization}:\\
		(4) compute $\boldsymbol{D}$ according to (\ref{equ_keypara1}), using $\hat{\boldsymbol{H}}_d$ and $\hat{\boldsymbol{G}}^H_i$;\\
		(5) solve (\ref{equs_Opt3}) using CVX, and decompose $\boldsymbol{V}^{*} = \boldsymbol{U}\boldsymbol{\Sigma}\boldsymbol{U}^H$;\\
		(6) set $\boldsymbol{v}^{*}=\boldsymbol{U}\boldsymbol{\Sigma}^{1/2}\boldsymbol{e}$, and adjust $\theta_n^{*} = \arg (	\boldsymbol{v}_n^{*}/	\boldsymbol{v}^{*}_{N+1})$;\\
		\textbf{RIS-aided transmission:} During the remaining  coherence time, all IoTs jointly transmit their data to the center buoy. 
	\end{algorithmic} \vspace{-0.1cm}
\end{algorithm}

\vspace{-0.1cm}

\subsection{RIS Optimization}

Given estimated $\boldsymbol{h}^d_i$ and $\boldsymbol{G}_i$ for each IoT $i$, the objective function $\sum_{i=1}^I ||(\boldsymbol{h}_i^d + \boldsymbol{q} \boldsymbol{G}_i)s_i||^2$ in (\ref{equ_obj}) can rewritten as : \small
\begin{multline} 
	\sum_{i=1}^I P_{t,i} \left[  ||\boldsymbol{h}_i^d ||^2 + \boldsymbol{q} \boldsymbol{G}_i (\boldsymbol{h}_i^d)^H  +  \boldsymbol{h}_i^d (\boldsymbol{q} \boldsymbol{G}_i)^H + \boldsymbol{q} \boldsymbol{G}_i(\boldsymbol{q} \boldsymbol{G}_i)^H \right]\\
	= \boldsymbol{q} \left( \sum_{i=1}^I P_{t,i} \boldsymbol{G}_i  \boldsymbol{G}_i^H \right) \boldsymbol{q}^H + \boldsymbol{q} \left(\sum_{i=1}^I P_{t,i}  \boldsymbol{G}_i (\boldsymbol{h}_i^d)^H\right)  + \\
	\left( \sum_{i=1}^I P_{t,i} \boldsymbol{h}_i^d \boldsymbol{G}_i^H \right) \boldsymbol{q}^H + \sum_{i=1}^I P_{t,i}  ||\boldsymbol{h}_i^d ||^2
\end{multline}\normalsize
Defining $\boldsymbol{v}=[\boldsymbol{q}, t]$, where $t$ is an auxiliary variable, and 
\begin{equation} \label{equ_keypara1}
	\boldsymbol{D}=
	\begin{bmatrix}
		 \sum_{i=1}^I P_{t,i} \boldsymbol{G}_i  \boldsymbol{G}_i^H &  \sum_{i=1}^I P_{t,i}  \boldsymbol{G}_i (\boldsymbol{h}_i^d)^H\\
		\sum_{i=1}^I P_{t,i} \boldsymbol{h}_i^d \boldsymbol{G}_i^H & \sum_{i=1}^I P_{t,i}  ||\boldsymbol{h}_i^d ||^2
	\end{bmatrix}. 
\end{equation}
The optimization problem (\ref{equs_Opt}) then can be homogenized as:
\begin{subequations}\label{equs_Opt2}  
	\begin{align}
		\max_{ \boldsymbol{v}=[q_1,\cdots, q_N,t]  } \quad &  \boldsymbol{v} \boldsymbol{D} \boldsymbol{v}^H   \label{equ_obj2}\\ 
		\textrm{s. t.} \quad  
		& |q_n|^2 =1,  \forall n = 1,\cdots, N,\\
		& |t|^2 = 1. 
	\end{align}
\end{subequations} 
Let $\boldsymbol{V}=\boldsymbol{q}^H\boldsymbol{q}$, then $\boldsymbol{q} \boldsymbol{D} \boldsymbol{q}^H = \text{Tr}(\boldsymbol{D}\boldsymbol{V})$, where Tr$(\cdot)$ denotes the matrix trace. 
Thus, (\ref{equs_Opt2}) is  reformulated as  \cite{jiang2023capacity}: 
\begin{subequations}\label{equs_Opt3}  
	\begin{align}
		\max_{ \boldsymbol{V} } \quad &  \text{Tr}(\boldsymbol{D}\boldsymbol{V}) \\ 
		\textrm{s. t.} \quad  
		& \boldsymbol{V}_{n,n}=1,  \forall n = 1,\cdots, N+1 \\
		& \boldsymbol{V} \succ 1, \label{cons_psdm}
	\end{align}
\end{subequations}
where $\boldsymbol{V}_{n,n}$ is the $n$-th diagonal element of $\boldsymbol{V}$, and (\ref{cons_psdm})  requires  $\boldsymbol{V}$ to be a   positive semi-definite matrix. 
Consequently, the optimization (\ref{equs_Opt3}) is transformed to a semi-definite program, whose globally optimal $\boldsymbol{V}^{*}$ can be efficiently found via numerical algorithms such as CVX. 

To get the optimal RIS coefficient $\boldsymbol{q}^{*}$,   the eigenvalue decomposition is conducted to $\boldsymbol{V}^{*} = \boldsymbol{U}\boldsymbol{\Sigma}\boldsymbol{U}^H$.  
Then, a sub-optimal solution to the optimization problem in (\ref{equs_Opt2}) is 
\begin{equation}
	\boldsymbol{v}^{*}=\boldsymbol{U}\boldsymbol{\Sigma}^{1/2}\boldsymbol{e},
\end{equation}
where $\boldsymbol{e} \sim \mathcal{CN}(\boldsymbol{0},\boldsymbol{I}_{N+1})$ is a Gaussian random vector. 
And the solution to  (\ref{equs_Opt}) can be determined as
\begin{equation}
	\theta_n^{*} = \arg (	\boldsymbol{v}_n^{*}/	\boldsymbol{v}^{*}_{N+1}), 
\end{equation}
with the maximized capacity of RIS-aided communication as 
\begin{multline}\vspace{-0.1cm}
	C^{*} = \beta \log_2  \left[1 + \sum_{i=1}^I P_{t,i} ||(\boldsymbol{h}_i^d + \boldsymbol{q }^{*} \boldsymbol{G}_i)||^2/\sigma^2 \right] \\
	= \beta \log_2  \left[1 + \sum_{i=1}^I P_{t,i} \left|\sum_{n=1}^{N} |\boldsymbol{h}_i^d| + |\boldsymbol{g}_{n,i}|\right|^2/\sigma^2  \right], 
\end{multline}
where $\boldsymbol{q }^{*} = [e^{j\theta^*_1}, \cdots,e^{j\theta^*_N}]$ and $\boldsymbol{G}_i = [\boldsymbol{g}^H_{1,i}, \cdots, \boldsymbol{g}^H_{N,i} ]^H$. 
The optimized  transmission for RIS-aided MU-MIMO maritime IoT systems is summarized in Algorithm 1, with the computational complexity of $\mathcal{O}(N^4 + IN^2M + IMN^2 + I^3 + I^2M)$.  
The impact of CSI errors and imperfect synchronization among multiple IoT transmissions is left for  future work.

\vspace{-0.1cm}

\section{Simulation Results and Analysis}

\begin{figure*}[!t]
	\begin{center}\vspace{-0.0cm}
		\begin{subfigure}{.32\textwidth}
			\centering
			\includegraphics[width=6.5cm]{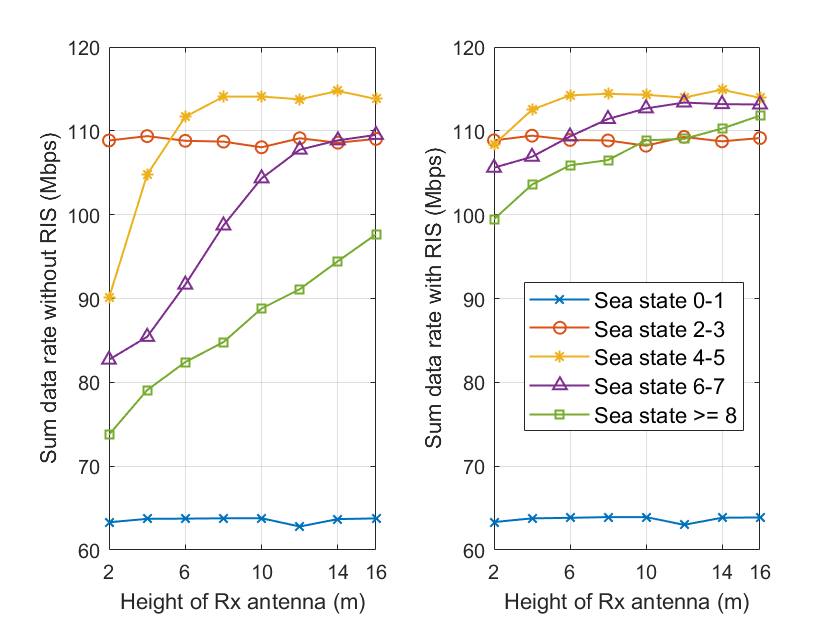}  
			\caption{\label{rate_hr} }
		\end{subfigure}
		\begin{subfigure}{.32\textwidth}
			\centering
			\includegraphics[width=6.4cm]{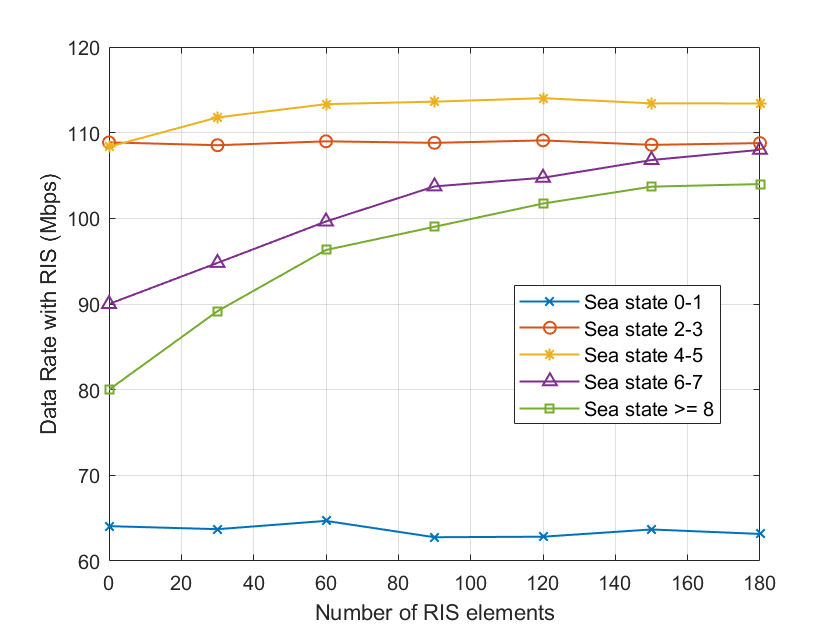}    
			\caption{\label{rate_numRIS}  }
		\end{subfigure}
		\begin{subfigure}{.32\textwidth}
			\centering
			\includegraphics[width=6.5cm]{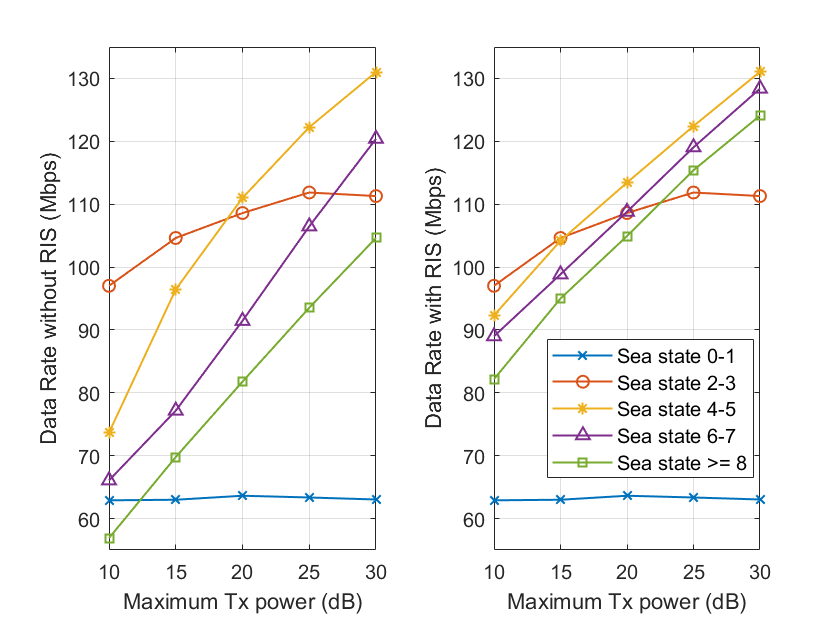}  
			\caption{\label{rate_power} }
		\end{subfigure}
		\vspace{-0.2cm}
		\caption{\small{\label{fig_sim} Sum of IoT transmission rates with and without the RIS, under different sea states, given an increasing (a)  antenna height $h_r^0$ of the center buoy (Rx), (b) number of RIS elements $N$, and (c)  maximum transmit powers $P_{\text{max}}$. }
		}
	\end{center}
	\vspace{-0.7cm}
\end{figure*}

In our simulations, the RIS is mounted onto the surface of an offshore wind turbine, with a diameter of $6$ meters and a height range of $20$ to $50$ meters.  
The antenna heights of Tx (IoT device) and Rx (center node) relative to their respective buoys are $h_t^0 = 2$ meters and $h_r^0 = 5$ meters. 
The evaporation duct height is $h_e = 50$ meters. 
The carrier frequency is $f_c = 5.8$ GHz with a bandwidth of $\beta =5$ MHz. 
The small-scale fading parameters are $\sigma_{LoS} = 3.5$ dB and $\sigma_{LoS}^2 = 5.1$ dB. 
The antenna gains are $G_t=0$ dB and $G_r=5$ dB. 
For the NLos path loss model, we adopt $K= 130.6$ dB and path loss exponent $\alpha =2.1$ dB from \cite{reyes2011buoy}. 
For the EH system, the PTO efficiency is $\eta_{\text{pto}} = 50\%$, the electronics conversion efficiency is $\eta_{\text{conv}}=90\%$, and CWR $\gamma_{\text{cwr}} \approx 8.2\%$ 
for point absorber WEC with with an effective interaction width of $W=  2$ meter \cite{babarit2015database}.  
For the communication system,  we consider  $N=360$ RIS elements, $M=8$ antennas at the center buoy,  and a maximum transmit power $P_{\text{max}}= 20$ dBW. 
The center buoy is located $200$ meters away from the wind turbine, with an average number of $I=4$ IoT buoys randomly deployed  within a $200$-meter radius of the wind turbine following a spatial Poisson process. 
The channel coefficient follows a block-fading model, which remains constant within each coherence interval and varies independently across intervals, and the buoy and RIS positions shift between intervals due to sea waves.

In Fig. \ref{fig_sim}, we compare the sum data rates of all IoT devices with and without RIS assistance. 
As shown in Fig. \ref{rate_hr}, the sum-rate increases with the receiver antenna height $h_r^0$, under sea state conditions $\ge 4$.  
In calm conditions (sea state $<4$),  wave height is low, thus, LoS links dominants the channel between IoT devices and the receiver. 
However,  limited wave activity also leads to insufficient harvested energy, making the system power-constrained and limiting the benefits of RIS.    
As sea states intensify ($\ge 4$), higher wave heights increase the NLoS probability, making the data rate more sensitive to  $h_r^0$. 
Thus,  a taller receiver antenna improves the LoS probability, as well as the overall transmission rate. 
In these conditions, RIS also provides substantial benefits, especially when the receiver antenna is low, by compensating for blocked direct links. 
Fig. \ref{los_hr} has shown that for sea state $4-5$, the LoS probability reaches nearly $100\%$ when $h_r^0 > 10$ m. Thus, beyond this height, RIS offers little  gain for these sea states. 
However, for sea states $\ge 6$, both increased antenna height and RIS assistance continue to improve the data rate performance.

Next, we evaluate the impact of RIS size on the transmission rate. As shown in Fig. \ref{rate_numRIS}, RIS has minimal impact in calm conditions (sea state $<4$) due to power limitations. 
In contrast, under rough sea conditions (sea state $\ge 4$), the data rate increases with the number of deployed RIS elements.
However, when the elements number exceeds $120$, the performance gain becomes marginal. 
This is due to increased overhead for channel estimation, which reduces the effective time available for data transmission. Hence, deploying a moderate number of RIS elements is more practical in maritime IoT systems.  

Finally, Fig. \ref{rate_power} examines the sensitivity of system performance to maximum transmit power $P_{\text{max}}$ under different sea states.  
In calm seas, limited harvested energy constrains transmit power, thus, increased $P_{\text{max}}$ or RIS deployment does not work effectively. 
In contrast, under rough sea conditions, where wave energy harvesting is sufficient, a higher allowed transmit power $P_{\text{max}}$ substantially enhances the data rate. 
In these scenarios, RIS further improves performance, particularly when the receiver antenna is low and direct paths are frequently blocked.

\section{Conclusion}

In this paper, we have proposed a novel RIS-assisted  framework to improve the maritime IoT communications by leveraging offshore infrastructures to enhance signal coverage and reliability.  
A sea wave–aware channel model and wave EH system have been developed to support self-powered IoT operations. 
Real-time CSI measurement has been designed to enable dynamic RIS optimization for improved multi-user communication.  
Simulation results show that the proposed system significantly improves IoT communication performance under harsh sea conditions, particularly when LoS links are frequently blocked. To our knowledge, this is the first work to integrate RIS and mechanical design for resilient maritime IoT networks.

\bibliographystyle{IEEEtran}
\bibliography{references}

\begin{thebibliography}{10}
\providecommand{\url}[1]{#1}
\csname url@samestyle\endcsname
\providecommand{\newblock}{\relax}
\providecommand{\bibinfo}[2]{#2}
\providecommand{\BIBentrySTDinterwordspacing}{\spaceskip=0pt\relax}
\providecommand{\BIBentryALTinterwordstretchfactor}{4}
\providecommand{\BIBentryALTinterwordspacing}{\spaceskip=\fontdimen2\font plus
\BIBentryALTinterwordstretchfactor\fontdimen3\font minus
  \fontdimen4\font\relax}
\providecommand{\BIBforeignlanguage}[2]{{%
\expandafter\ifx\csname l@#1\endcsname\relax
\typeout{** WARNING: IEEEtran.bst: No hyphenation pattern has been}%
\typeout{** loaded for the language `#1'. Using the pattern for}%
\typeout{** the default language instead.}%
\else
\language=\csname l@#1\endcsname
\fi
#2}}
\providecommand{\BIBdecl}{\relax}
\BIBdecl

\bibitem{huo2020cellular}
Y.~Huo, X.~Dong, and S.~Beatty, ``Cellular communications in ocean waves for
  maritime {I}nternet of {T}hings,'' \emph{IEEE Internet of Things Journal},
  vol.~7, no.~10, pp. 9965--9979, 2020.

\bibitem{wang2018wireless}
J.~Wang, H.~Zhou, Y.~Li, Q.~Sun, Y.~Wu, S.~Jin, T.~Q. Quek, and C.~Xu,
  ``Wireless channel models for maritime communications,'' \emph{IEEE access},
  vol.~6, pp. 68\,070--68\,088, 2018.

\bibitem{alqurashi2022maritime}
F.~S. Alqurashi, A.~Trichili, N.~Saeed, B.~S. Ooi, and M.-S. Alouini,
  ``Maritime communications: A survey on enabling technologies, opportunities,
  and challenges,'' \emph{IEEE Internet of Things Journal}, vol.~10, no.~4, pp.
  3525--3547, 2022.

\bibitem{wang2021hybrid}
Y.~Wang, W.~Feng, J.~Wang, and T.~Q. Quek, ``Hybrid satellite-uav-terrestrial
  networks for 6{G} ubiquitous coverage: A maritime communications
  perspective,'' \emph{IEEE Journal on Selected Areas in Communications},
  vol.~39, no.~11, pp. 3475--3490, 2021.

\bibitem{reyes20145}
J.~Reyes-Guerrero and L.~Mariscal, ``5.8 {G}hz propagation of low-height
  wireless links in sea port scenario,'' \emph{Electronics Letters}, vol.~50,
  no.~9, pp. 710--712, 2014.

\bibitem{reyes2011buoy}
J.~Reyes-Guerrero, M.~Bruno, L.~A. Mariscal, and A.~Medouri, ``Buoy-to-ship
  experimental measurements over sea at 5.8 {G}hz near urban environments,'' in
  \emph{2011 11th Mediterranean Microwave Symposium (MMS)}.\hskip 1em plus
  0.5em minus 0.4em\relax IEEE, 2011, pp. 320--324.

\bibitem{wang2011effective}
H.~Wang, W.~Jia, and G.~Min, ``Effective channel exploitation in {IEEE} 802.16
  j networks for maritime communications,'' in \emph{2011 31st International
  Conference on Distributed Computing Systems}.\hskip 1em plus 0.5em minus
  0.4em\relax IEEE, 2011, pp. 162--171.

\bibitem{mi2025multi}
J.~Mi, J.~Huang, L.~Yang, A.~Ahmed, X.~Li, X.~Wu, R.~Datla, B.~Staby, M.~Hajj,
  and L.~Zuo, ``Multi-scale concurrent design of a 100 kw wave energy
  converter,'' \emph{Renewable Energy}, vol. 238, p. 121835, 2025.

\bibitem{babarit2015database}
A.~Babarit, ``A database of capture width ratio of wave energy converters,''
  \emph{Renewable Energy}, vol.~80, pp. 610--628, 2015.

\bibitem{zhang2019reflections}
Q.~Zhang, W.~Saad, and M.~Bennis, ``Reflections in the sky: Millimeter wave
  communication with {UAV}-carried intelligent reflectors,'' in \emph{Proc. of
  IEEE Global Communications Conference}, Hawaii, USA, Dec 2019.

\bibitem{chen2023channel}
J.~Chen, Y.-C. Liang, H.~V. Cheng, and W.~Yu, ``Channel estimation for
  reconfigurable intelligent surface aided multi-user mmwave {MIMO} systems,''
  \emph{IEEE Transactions on Wireless Communications}, vol.~22, no.~10, pp.
  6853--6869, 2023.

\bibitem{jiang2023capacity}
W.~Jiang and H.~D. Schotten, ``Capacity analysis and rate maximization design
  in {RIS}-aided uplink multi-user {MIMO},'' in \emph{IEEE Wireless
  Communications and Networking Conference (WCNC)}.\hskip 1em plus 0.5em minus
  0.4em\relax IEEE, 2023.

\end{thebibliography}

\end{document}